\documentclass{iopart}
 \usepackage{graphicx}
 \usepackage{bm}

\begin{document}

 \title{Enhanced van der Waals interaction at interfaces}
 \author{Marin-Slobodan Toma\v s}
 \address{Rudjer Bo\v skovi\' c Institute, P. O. B. 180,
 10002 Zagreb, Croatia}
 \ead{tomas@thphys.irb.hr}

 \begin{abstract}
Using a recently obtained (general) formula for the interaction
energy between an excited and a ground-state atom (Sherkunov Y
2007 {\it Phys. Rev.} A {\bf 75} 012705), we consider the
interaction energy between two such atoms near the interface
between two media. We demonstrate that under the circumstances of
the resonant coupling of the excited atom to the surface polariton
mode of a vacuum-medium system the nonretarded atom*-atom
interaction energy can be enhanced by (several) orders of
magnitude in comparison with the van der Waals interaction energy
of the two isolated atoms.
\end{abstract}

\section{Introduction}
It is known for quite some time that the decay rate and energy of
an excited atom (atom*) may be strongly modified near a dispersive
and absorbing surface owing to the resonant coupling of the atom
to the surface polariton mode(s) of the system
\cite{CPS,WS1,WS2,Bar,Fich,Buh,GoDu}. Excitation of surface
polaritons by decaying atom (or molecule) and their subsequent
conversion into the radiation (photons) by some means is in the
core of surface enhanced optical processes such as fluorescence
and Raman scattering \cite{CF,FW}. Similarly, owing to the
resonant energy shift of the atom, a strong modification of the
related atom*-surface van der Waals force leading even to the
atom-surface repulsion has been predicted
\cite{CPS,WS2,Fich,Buh,GoDu} and observed \cite{Fail}. In this
work we demonstrate yet another spectacular consequence of the
resonant atom*-surface coupling, namely the possibility of a
strong enhancement of the van der Waals interaction between an
excited and a ground-state atom in vicinity of an interface that
supports surface modes.

The atom*-atom interaction has so far been considered only for
atoms in free-space \cite{LP,GLP,Phil,PT1,PT2,Rizz} and lately for
atoms embedded in an absorbing medium \cite{BN,YS}. A
straightforward way to study this interaction in an inhomogeneous
system would therefore be to use a macroscopic QED approach
appropriate for absorbing systems and derive the interaction
potential between the atoms, for example, along the same lines as
it was recently done for two ground-state atoms \cite{Saf}.
However, instead of developing the theory from the beginning, in
this work we adopt the Sherkunov formula for the atom*-atom
interaction potential in an absorbing medium \cite{YS}. Indeed,
this formula is given in terms of the corresponding Green function
and there is nothing in its derivation which crucially depends on
the specific system considered in the paper. Accordingly, with the
appropriate Green function, it can be applied to an inhomogeneous
medium as well.

\section{Theory}
Consider an excited ($e$) atom A and a ground-state ($g$) atom B
embedded in an inhomogeneous magnetoelectric system described by
the permittivity $\varepsilon({\bf r},\omega)$ and permeability
$\mu({\bf r},\omega)$. Assuming, for simplicity, two-level
isotropic atoms, the vacuum force on the atom A can then be
obtained from the potential
\begin{equation}
\label{UA} U_A({\bf r}_A,{\bf r}_B)=U_A({\bf r}_A)+ U _{AB}({\bf
r}_A,{\bf r}_B),
\end{equation}
where
\begin{equation}
\label{UAm} \fl U_{A}({\bf r}_A)= \frac{\hbar}{2\pi c^2}
\int_0^\infty \rmd\xi\xi^2\alpha^A_e(\rmi\xi) \Tr\left[\mathbf
{G}^{\rm sc}({\bf r}_A,{\bf r}_A;\rmi\xi)\right]
 -\frac{|{\bf d}^A_{eg}|^2\omega_A^2}{3c^2}{\rm Re}
\Tr\left[\mathbf{G}^{\rm sc}({\bf r}_A,{\bf r}_A;\omega_A)\right]
\end{equation}
is the Casimir-Polder potential due to the inhomogeneity of the
system \cite{WS2,Buh,GoDu} and
\begin{eqnarray}
\label{UAB} \fl U_{AB}({\bf r}_A,{\bf r}_B)&= -\frac{\hbar}{2\pi
c^4}\int_0^\infty
\rmd\xi\xi^4\alpha^A_e(\rmi\xi)\alpha^B_g(\rmi\xi)
\Tr\left[\mathbf{G}({\bf r}_A,{\bf r}_B;\rmi\xi)\cdot
\mathbf{G}({\bf r}_B,{\bf r}_A;\rmi\xi)\right]\nonumber\\
&-\frac{{\rm Re}[\alpha^B_g(\omega_A)]\omega^4_A}{3c^4}|{\bf
d}^A_{eg}|^2\Tr\left[\mathbf{G}({\bf r}_A,{\bf
r}_B;\omega_A)\cdot\mathbf{G}^*({\bf r}_B,{\bf
r}_A;\omega_A)\right]
\end{eqnarray}
is the van der Waals (interaction) potential between the atoms A
and B [Ref. \cite{YS}, equation (68)]. Here
\begin{equation}
\label{alpha} \alpha^X_{e(g)}(\omega)=-(+)\frac{2|{\bf
d}^X_{eg}|^2} {3\hbar\omega_X}\frac{\omega_X^2}
{\omega_X^2-\omega^2-\rmi\omega\gamma_X},\;\;\;X=A\;{\rm or}\;B,
\end{equation}
are the atomic polarizabilities, $\omega_X=(E^X_e-E^X_g)/\hbar$
and ${\bf d}^X_{eg}=<e|{\bf d}^X|g>$ are the transition frequency
and the dipole matrix element of atom $X$, respectively, whereas
$\gamma_X$ is the width of its excited state [$\gamma_A=0^+$ in
(\ref{UAm}) and (\ref{UAB}) owing to the approaches adopted when
deriving these results]. The dyadic
\begin{equation}
\label{G0G}
\mathbf{G}({\bf r},{\bf r'};\omega;)=\mathbf{G}^0({\bf
r},{\bf r'};\omega)+\mathbf{G}^{\rm sc}({\bf r},{\bf r'};\omega),
\end{equation}
with $\mathbf{G}^0({\bf r},{\bf r'};\omega)$ being the Green
function in a homogeneous medium, is the classical Green function
for the system satisfying [$\mathbf{I}=\hat{\bf x}\hat{\bf
x}+\hat{\bf y}\hat{\bf y}+\hat{\bf z}\hat{\bf z}$]
 \begin{equation}
 \left[\nabla\times\frac{1}{\mu({\bf r},\omega)}\nabla\times-
 \varepsilon({\bf r},\omega)\frac{\omega^2}{c^2}\mathbf{I}\cdot\right]
 \mathbf{G}(\omega;{\bf r},{\bf r'})
 =4\pi\mathbf{I}\delta({\bf r}-{\bf r'}),
 \label{GF}
 \end{equation}
 with the outgoing wave condition at infinity.

All information about the mode structure of the system are
contained in the scattering part of the Green function
$\mathbf{G}^{\rm sc}({\bf r},{\bf r'};\omega)$. Specially, poles
of this Green function in the complex $\omega$-plane correspond to
frequencies of the system resonant (polariton) modes. Accordingly,
whenever $\omega_A$ is close to the frequency $\omega_r$ of a
resonant mode $U_A({\bf r}_A,{\bf r}_B)$ is strongly modified
owing to the second (resonant) terms in (\ref{UAm}) and
(\ref{UAB}). Since these terms are absent if the atom A is in the
ground state this effect appears only for excited atoms.
Evidently, when $\omega_A\sim\omega_r$, the Casimir-Polder
potential $U_A({\bf r}_A)$ exhibits a dispersion due to the atom
coupling to the resonant mode. Simultaneously, in addition of
showing the intrinsic (genuine) dispersive type resonance for
$\omega_A\sim\omega_B$ governed by ${\rm
Re}[\alpha^B_g(\omega_A)]$, the van der Waals potential is
resonantly enhanced when $\omega_A\sim\omega_r$ owing to the
factor $|G_{ij}({\bf r}_A,{\bf r}_B;\omega_A)|^2$ in (\ref{UAB}).
From this factor, we may infer that the resonant enhancement of
$U_{AB}({\bf r}_A,{\bf r}_B)$ with respect to its free-space value
is due to the exchange of (real) system excitations between the
atoms (instead of ordinary photons, as in the free space
\cite{PT1,PT2}) and the associated field-intensity enhancement at
atomic sites \cite{CF,FW}.

\begin{figure}[htb]
\begin{center}
\resizebox{8cm}{!}{\includegraphics{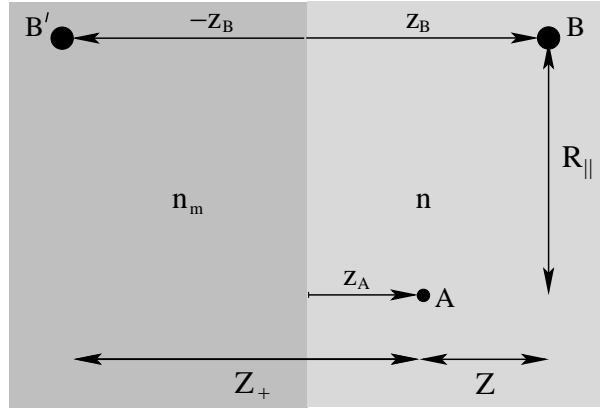}}
\end{center}
 \caption{Two atoms near an interface shown schematically
 (atom B' is the mirror image of the atom B).
 Media are described by (complex) refraction indexes
 $n(\omega)=\sqrt{\varepsilon(\omega)\mu(\omega)}$ and
 $n_m(\omega)=\sqrt{\varepsilon_m(\omega)\mu_m(\omega)}$.
\label{AB}}
\end{figure}

To illustrate the above considerations and estimate the
enhancement of the atom*-atom van der Waals interaction, we assume
that the atoms $A$ and $B$ are embedded in a system consisting of
two semi-infinite media, as depicted in Fig. \ref{AB}. To account
for the local-field effects, we adopt the Onsager model and
therefore also assume small empty spherical cavities around the
atoms. Provided that $\omega_{\rm max} R_X/c\ll 1$, with $R_X$
being cavity radii and $\omega_{\rm max}>\omega_X$ an effective
cutoff frequency in (\ref{UAm}) and (\ref{UAB}), the Green
function for this system can (to the order of $\omega R_X/c$) be
written as \cite{Sam}
 \numparts
\begin{equation}
\label{Glsc} \mathbf{G}^{\rm sc}({\bf r}_A,{\bf
r}_A;\omega)=i\frac{2\omega}{3c}C(\omega)\mathbf{I}+
L(\omega)\tilde{\mathbf{G}}^{\rm sc}({\bf r}_A,{\bf r}_A;\omega),
\end{equation}
\begin{equation}
\label{Gl} \mathbf{G}({\bf r}_A,{\bf r}_B;\omega)=
L(\omega)\tilde{\mathbf{G}}({\bf r}_A,{\bf r}_B;\omega),\;\;\;
L(\omega)=\left[\frac{3\varepsilon(\omega)}{2\varepsilon(\omega)+1}\right]^2.
\end{equation}
\endnumparts
Here $C(\omega)$ (given explicitly in Ref. \cite{Sam}) is the
reflection coefficient for the field scattered within the cavity
and $\tilde{\mathbf{G}}({\bf r}_A,{\bf r}_B;\omega)$ is the Green
function for the system unperturbed by the Onsager cavities. In
order to keep the discussion simple, in this work we consider the
situation where the atomic distances from the interface between
the media are small compared to $c/\omega_{\rm max}$ (but larger
than $R_X$) so that the retardation of the electromagnetic field
can be neglected. The Green function for this system in the
nonretarded approximation $\tilde{\mathbf{G}}_{\rm nr}({\bf
r}_A,{\bf r}_B;\omega)$ can be found by adding the nonretarded
(quasistatic) field ${\bf E}_{\rm nr}({\bf r}_A,{\bf r}_B;\omega)$
of an oscillating dipole ${\bf d}$ at ${\bf r}_B$ and the
corresponding field ${\bf E}'_{\rm nr}({\bf r}_A,{\bf
r}'_B;\omega)$ of its image
\begin{equation}\label{r}
{\bf d}'=r(\omega)(-\mathbf{I}_\parallel+\hat{\bf z}\hat{\bf
z})\cdot{\bf d},\;\;\;\;\;
r(\omega)=\frac{\varepsilon_m(\omega)-\varepsilon(\omega)}
{\varepsilon_m(\omega)+\varepsilon(\omega)},
\end{equation}
at ${\bf r}'_B={\bf r}_{B\parallel}-z_B\hat{\bf z}$. Here
$\mathbf{I}_\parallel=\hat{\bf x}\hat{\bf x}+\hat{\bf y}\hat{\bf
y}$ and $r(\omega)$ is the Fresnel reflection coefficient for the
interface in the quasistatic limit. Thus, using \cite{Tom95}
\begin{equation}
{\bf E}^{\rm tot}_{\rm nr}({\bf r}_A,{\bf r}_B;\omega)
=\frac{\omega^2}{c^2}\tilde{\mathbf{G}}_{\rm nr}({\bf r}_A,{\bf
r}_B;\omega)\cdot{\bf d}
\end{equation}
and Eq. (\ref{Gl}), we have for the Green function
\begin{equation}
\label{Gnr} \fl \mathbf{G}_{\rm nr}({\bf r}_A,{\bf r}_B;\omega)=
\frac{c^2L(\omega)}{\varepsilon(\omega)\omega^2}\left[\frac{3{\bf
R}{\bf R}-\mathbf{I}R^2}{R^5} +r(\omega) \frac{(3{\bf R}'{\bf
R}'-\mathbf{I}R'^2) \cdot(-\mathbf{I}_\parallel+ \hat{\bf
z}\hat{\bf z})}{R'^5} \right],
\end{equation}
where ${\bf R}={\bf r}_A-{\bf r}_B={\bf R}_\parallel+Z\hat{\bf z}$
and ${\bf R}'={\bf r}_A-{\bf r}'_B={\bf R}_\parallel+Z_+\hat{\bf
z}$, with $Z=z_A-z_B$ and $Z_+=z_A+z_B$.

Combining the nonretarded limit \cite{Tom95}
of (\ref{G0G}) and (\ref{Glsc}) with (\ref{Gnr}), we find that
\begin{equation}
\Tr[\mathbf{G}^{\rm sc}_{\rm nr}({\bf r}_A,{\bf r}_A;\omega)]
=\frac{6c^2}{\omega^2R_A^3}\frac{\varepsilon(\omega)-1}
{2\varepsilon(\omega)+1}+ \frac{c^2}{2\omega^2z_A^3}
\frac{L(\omega)r(\omega)}{\varepsilon(\omega)}.
\end{equation}
In conjunction with (\ref{UAm}), this leads to (we omit the
position-independent part of the potential due to the nearby
medium)
\begin{equation}\fl
\label{UAs} U_{A}({\bf r}_A)=-\frac{\hbar}{4\pi
z_A^3}\int_0^\infty
\rmd\xi\alpha^A_e(\rmi\xi)\frac{L(\rmi\xi)r(\rmi\xi)}
{\varepsilon(\rmi\xi)}
 -\frac{|{\bf d}^A_{eg}|^2}{6z_A^3} {\rm
Re}\frac{L(\omega_A)r(\omega_A)}{\varepsilon(\omega_A)},
\end{equation}
which generalizes a well-known formula for the Casimir-Polder
potential of an excited atom near an interface \cite{WS2} by
including the effect of the surrounding medium.

Similarly, inserting $\mathbf{G}_{\rm nr}({\bf r}_A,{\bf
r}_B;\omega)$ in (\ref{UAB}), we obtain
\begin{eqnarray}
\label{UABs} \fl U_{AB}({\bf r}_A,{\bf r}_B)&=
-\frac{\hbar}{\pi}\int_0^\infty
\rmd\xi\alpha^A_e(\rmi\xi)\alpha^B_g(\rmi\xi)
\frac{L^2(i\xi)}{\varepsilon^2(\rmi\xi)}
W(R_\parallel,Z,Z_+;\rmi\xi)
\nonumber\\
&-\frac{2|{\bf d}^A_{eg}|^2\left|L(\omega_A)\right|^2}
{3|\varepsilon(\omega_A)|^2} {\rm Re}[\alpha^B_g(\omega_A)]
W(R_\parallel,Z,Z_+;\omega_A),
\nonumber\\
\fl
W(R_\parallel,Z,Z_+;\omega)&=\frac{3}{R^6}+|r^2(\omega)|^2\frac{3}{R'^6}
-{\rm Re}[r(\omega)]\frac{3(R^4_\parallel-Z^2Z_+^2)+
R^2R'^2}{R^5R'^5}.
\end{eqnarray}
Here, the first two terms come from the direct interaction between
the atoms and the interaction of atom A with the image of atom B,
respectively, whereas the third one is an interference term.

According to (\ref{alpha}), the off-resonant (first) term in
(\ref{UABs}) is of the same form as the van der Waals potential
between two ground-state atoms. Therefore, referring the reader to
Ref. \cite{Saf} for a detailed discussion of this term, here we
pay attention only to the (usually much larger) resonant part of
the atom*-atom potential $U^{\rm r}_{AB}$. Using (\ref{alpha}), we
rewrite $U^{\rm r}_{AB}$ as
\begin{eqnarray}
\label{UABr} \fl U_{AB}^{\rm r}({\bf r}_A,{\bf
r}_B)=&-\frac{2|{\bf d}^A_{eg}|^2\left|L(\omega_A)\right|^2}
{|\varepsilon(\omega_A)|^2}\frac{\alpha^B_g(0)}{R^6}
\frac{\omega_B^2(\omega_B^2-\omega_A^2)}{(\omega_B^2-\omega_A^2)^2
+(\omega_A\gamma_B)^2}\nonumber\\
&\times\left[1+|r^2(\omega_A)|^2\frac{R^6}{R'^6} -\frac{1}{3}{\rm
Re}[r(\omega_A)]\frac{R}{R'} \frac{3(R^4_\parallel-Z^2Z_+^2)+
R^2R'^2}{R'^4}\right].
\end{eqnarray}
Evidently, the last factor here describe the modification of the
resonant atom*-atom potential in an infinite medium owing to the
presence of the nearby interface. As seen, it is highly
anisotropic and depends not only on the distance of the molecules
from the interface but also on their mutual orientation with
respect to it.

\section{Discussion}
In order to keep the discussion simple, we assume that the atoms
are embedded in the vacuum [$\varepsilon(\omega)=1$] in front of a
dielectric medium that is around a resonance at $\omega_T$
described by the dielectric function \cite{Bar}
\begin{equation}
\varepsilon_m(\omega)=\eta\left(1+\frac{\omega_P^2}
{\omega_T^2-\omega^2-\rmi\omega\Gamma}\right),
\end{equation}
where $\eta$ is a background dielectric constant, $\eta\omega_P^2$
is (essentially) the coupling between the medium polarization and
the field and $\Gamma$ is the corresponding damping constant. We
note that this dielectric function is physically unacceptable
generally since it does not tend to unity for large frequencies.
However, it can safely be employed as an effective permittivity in
a (finite) frequency interval around $\omega_T$. The reflection
coefficient to be used in (\ref{UABr}) therefore reads
\begin{equation}
\label{rs} 
 r(\omega)=\frac{\eta-1}{\eta+1}+
 \sigma^2\frac{\omega_S^2}{\omega_S^2-\omega^2-\rmi\omega\Gamma},
 \;\;\sigma^2=\frac{\varepsilon(0)-1}{\varepsilon(0)+1}
-\frac{\eta-1}{\eta+1},
\end{equation}
where $\omega_S=\sqrt{\eta\omega_P^2/(\eta+1)+\omega_T^2}$ is the
surface-mode frequency.
Equations (\ref{UABr}) and (\ref{rs}) explicitly demonstrate the
intrinsic dispersion of $U_{AB}^{\rm r}$ when
$\omega_A\sim\omega_B$ and enhancement when
$\omega_A\sim\omega_S$. Note that for
$\omega_B>\omega_A\sim\omega_S$ ($\omega_B<\omega_A\sim\omega_S$)
this surface enhanced potential is attractive (repulsive). It is
also immediately seen that when $\omega_A=\omega_S$ the potential
$U_{AB}^{\rm r}$ is enhanced with respect to its free-space value
by a factor of
\begin{equation}
g(R_\parallel,z_A,z_B;\omega_S)\simeq \frac{\sigma^4\omega_S^2}
{\Gamma^2}\left(1+\frac{4z_Az_B} {R^2}\right)^{-3}.\label{g}
\end{equation}
Since for insulators typically $\Gamma/\omega_S\sim 10^{-2}$ and
for (noble) metals ($\eta=1$ and $\omega_T=0$) typically
$\Gamma/\omega_S\sim 10^{-3}$, this implies that under the
circumstances of resonant coupling of atom A to the surface
polariton mode at nearby surface $U_{AB}^{\rm r}$ could be
enhanced by several orders of magnitudes.

\begin{figure}[htb]
\begin{center}
\resizebox{8cm}{!}{\includegraphics{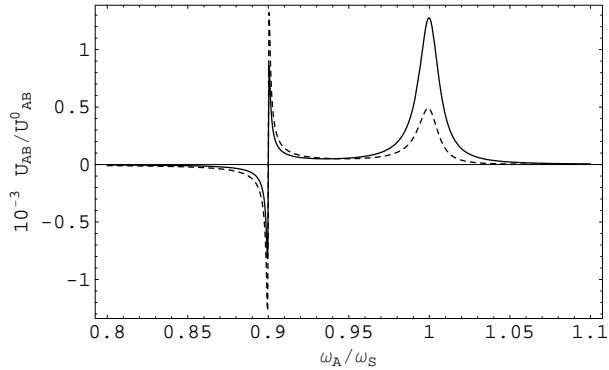}}
\end{center}
 \caption{Relative interaction potential for parallel (solid line)
 and perpendicular (dashed line) orientation of the atoms with respect
 to the surface at the distance $z_A=0.1R$ from the atom A as a function
 of the transition frequency of atom A.
 Relevant medium parameters are $\eta=2.71$,
 $\varepsilon(0)=6.57$, $\Gamma=0.015\omega_S$ and $\omega_S=
 1.54\times 10^{14}{\rm s}^{-1}$ \cite{GoDu}. The transition
 frequency and linewidth of
 atom B are $\omega_B=0.9\omega_S$ and $\gamma_B=10^{-3}\omega_S$,
 respectively. \label{Upo}}
\end{figure}
\begin{figure}[htb]
\begin{center}
\resizebox{8cm}{!}{\includegraphics{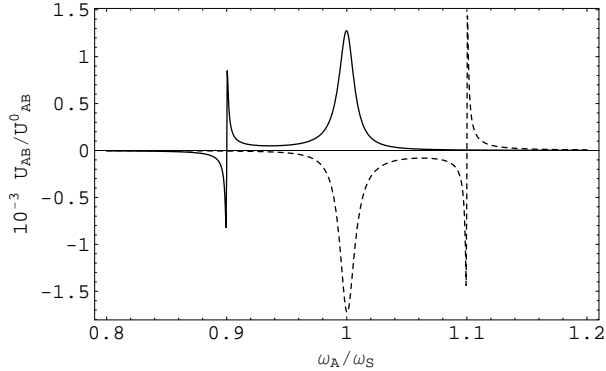}}
\end{center}
 \caption{Relative interaction potential for parallel orientation of
 the atoms and for $\omega_B=0.9\omega_S$ (solid line) and
 $\omega_B=1.1\omega_S$ (dashed line). Other parameters are the same as
 in Fig. \ref{Upo}. \label{Upgg}}
\end{figure}
We illustrate the above considerations in Fig. \ref{Upo} where we
have plotted $U_{AB}^{\rm r}$ in units $U_{AB}^0=2|{\bf
d}^A_{eg}|^2\alpha^B_g(0)/R^6$ as a function of $\omega_A$ for
parallel and perpendicular orientation of the atoms with respect
to the surface. For these orientations of the atoms $U_{AB}^{\rm
r}$ is a (monotonically decreasing) function of $z_A/R$ only and
displayed curves correspond to $z_A=0.1R$. Medium parameters are
chosen from Ref. \cite{GoDu} and correspond to sapphire around the
surface polariton resonance at $\lambda_S= 12.21 {\rm \mu m}$
whereas parameters of the atom B are chosen quite arbitrarily. We
see that, besides a resonant structure at $\omega_B$ as would
exist in the free-space, the potential exhibits also a resonance
at the surface mode frequency $\omega_S$ implying a strongly
surface enhanced interaction between the atoms. For example, we
find that $U_{AB}^{\rm r}$ in the parallel geometry is at the
surface mode resonance 298.5 times larger than in the free space.
We note that this is in a very good agreement with
$g(R_\parallel,z_A,z_A;\omega_S)$ calculated from (\ref{g}) for
$z_A=0.1R$.

\begin{figure}[htb]
\begin{center}
\resizebox{8cm}{!}{\includegraphics{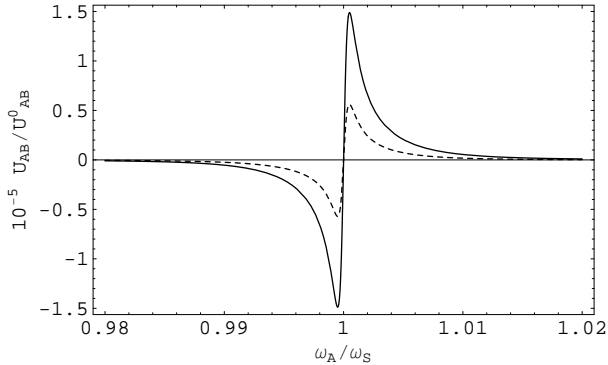}}
\end{center}
 \caption{Resonant interaction potential for parallel (solid line)
  and perpendicular (dashed line) orientation of the atoms for
  $\omega_A\sim\omega_B=\omega_S$. Other parameters are
  the same as in Fig. \ref{Upo}. \label{Udr}}
\end{figure}

Of course, for $\omega_B>\omega_S\sim\omega_A$, the surface
enhanced potential is attractive, as illustrated in Fig.
\ref{Upgg} where we have plotted $U_{AB}^{\rm r}$ in the parallel
geometry for two symmetric values of $\omega_B$ with respect to
$\omega_S$. The largest enhancement is, however, obtained when
$\omega_A\sim\omega_B\sim\omega_S$. This "double resonance" case
is illustrated in Fig. \ref{Udr}, where we have plotted
$U_{AB}^{\rm r}$ in the parallel (solid line) and perpendicular
(dashed line) geometry for $\omega_A$ around $\omega_B=\omega_S$.
We see that just below (above) the double resonance the attractive
(repulsive) part of the potential is additionally considerably
enhanced (note the change of scale on the ordinate axes).

We end this discussion by a short comment on the force on the atom
A
\begin{equation}
{\bf F}_A({\bf r}_A,{\bf r}_B)=-\nabla_AU_A({\bf r}_A,{\bf r}_B)
\end{equation}
around the surface mode resonance. Keeping only the (most)
resonant terms in (\ref{UAs}) and (\ref{UABs}), we find for ${\bf
F}_A$ [$\varepsilon(\omega)=1$]
\begin{equation}
\label{FA} \fl {\bf F}_A({\bf r}_A,{\bf r}_B)\simeq -\hat{\bf
z}\frac{|{\bf d}^A_{eg}|^2}{2z_A^4} {\rm
Re}[\underline{r}(\omega_A)]-{\bf R}'\frac{12|{\bf
d}^A_{eg}|^2}{R'^8} {\rm
Re}[\alpha^B_g(\omega_A)]|\underline{r}(\omega_A)|^2,
\end{equation}
where $\underline{r}(\omega)$ is given by the resonant term in
(\ref{rs}). Introducing the function
\begin{equation}
L(x,y,z)=x^4[(x^2-y^2)^2+(yz)^2]^{-1},
\end{equation}
we rewrite ${\bf F}_A$ at $\omega_A\simeq\omega_S$ in components
as
\numparts
\begin{equation}
\label{Fpar}
\fl {\bf F}_{A\parallel}({\bf r}_A,{\bf r}_B)\simeq
-{\bf R}_\parallel\frac{12|{\bf
d}^A_{eg}|^2\alpha^B_g(0)}{(R^2+4z_Az_B)^4}
(1-\frac{\omega_A^2}{\omega_B^2})
L(\omega_B,\omega_A,\gamma_B)\sigma^4L(\omega_S,\omega_A,\Gamma)
\end{equation}
\begin{eqnarray}
\label{FAz} \fl F_{Az}({\bf r}_A,{\bf r}_B)\simeq &-\frac{|{\bf
d}^A_{eg}|^2}{2z_A^4}\sigma^2
L(\omega_S,\omega_A,\Gamma)\left[1-\frac{\omega_A^2}{\omega_S^2}\right.
\nonumber\\
&\left.+ 24\frac{z_A^4(z_A+z_B)\alpha^B_g(0)}{(R^2+4z_Az_B)^4}
(1-\frac{\omega_A^2}{\omega_B^2}) L(\omega_B,\omega_A,\gamma_B)
\sigma^2\right].
\end{eqnarray}
\endnumparts
As seen, because of the presence of the atom B, ${\bf F}_A$
acquires a parallel component and its perpendicular component is
diminished or enlarged owing to the relative positions of the
atomic and surface resonances.

\begin{figure}[htb]
\begin{center}
\resizebox{8cm}{!}{\includegraphics{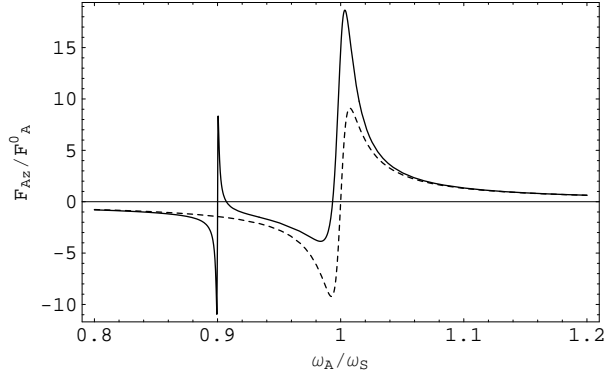}}
\end{center}
 \caption{Relative perpendicular force on atom A for parallel orientation
 of atoms A and B with respect to the sapphire surface. The distance of atom A
 from the surface is $z_A=3[\alpha^B_g(0)]^{1/3}$ and the distance between the atoms
is $R=z_A$ (solid line) and $R=5z_A$ (dashed line). Other
parameters are the same as in Fig. \ref{Upo}. \label{F}}
\end{figure}

The above considerations are illustrated in Fig. \ref{F} where we
have plotted $F_{Az}(\omega_A)$ in units $F^0_A=|{\bf
d}^A_{eg}|^2/2z_A^4$ for parallel orientation of atoms A and B
with respect to the surface and for two values of the parameter
$R/z_A$. The distance of atom A from the surface is fixed letting
$z_A=3[\alpha^B_g(0)]^{1/3}$; for $\alpha^B_g(0)$ in the range of
polarizabilities of the alkali-metal atoms [$\simeq (2-6)\times
10^{-23}\;{\rm cm}^3$ \cite{DeJo}], for example, this corresponds
to $z_A\sim 1$ nm. As follows from (\ref{FAz}), under these
circumstances the atom-atom component of the force decreases
rapidly with the relative distance between the atoms $R/z_A$ so
that the dashed line practically coincides with the purely
Casimir-Polder (atom-surface) force given by the first term in
(\ref{FAz}). When the atoms are closer (solid line), owing to its
strong repulsion from the atom B for $\omega_A\sim\omega_S$, the
force on atom A is diminished/enhanced in the attractive
($\omega_A<\omega_S$)/repulsive ($\omega_A>\omega_S$) atom-surface
force region. Of course, for $\omega_B>\omega_S$ the situation is
reversed: due to the attraction between the atoms around the
surface resonance $F_{Az}(\omega_A)$ is enhanced/diminished in the
attractive/repulsive atom-surface force region. The same effect is
also observed for the perpendicular orientation of atoms with
respect to the surface, however, owing to the attenuation od the
surface-mode field away from the surface, in this case it is much
weaker. Evidently, this effect is reinforced when the atoms are
closer to the surface and to each other and disappears for large
$z_A$ and/or $R$.

\section{Conclusions}
In conclusion, we have demonstrate that the (generalized)
Sherkunov formula implies a strong enhancement of the van der
Waals atom*-atom interaction near a vacuum-medium interface under
the circumstances of the resonant coupling of the excited atom
with the surface mode of the system. The enhancement is due to the
exchange of (real) surface excitations between the atoms (instead
of photons) and the accompanying enhancement of the
electromagnetic field intensity at atomic sites. This promotes the
van der Waals interaction as another example of the surface
enhanced phenomena.

\ack The author thanks Yuri Sherkunov for discussions and useful
comments. This work was supported by the Ministry of Science,
Education and Sport of the Republic of Croatia under contract No.
098-1191458-2870.

\end{document}